\documentclass[12pt]{iopart}
\usepackage{graphicx}
\usepackage{amsfonts}
\usepackage{amsthm}

\newcommand{\bB}{\mathbf{B}}
\newcommand{\bE}{\mathbf{E}}
\newcommand{\bF}{\mathbf{F}}

\newcommand{\cM}{\mathcal{M}}

\newcommand{\dmu}{\partial_\mu}

\newcommand{\Dmu}{\partial^\mu}

\begin{document}
\title{The scalar complex potential and the Aharonov-Bohm effect}
\author{Y. Friedman and V. Ostapenko}
\address{Jerusalem College of Technology \\
P.O.B. 16031 Jerusalem 91160, Israel}
\ead{friedman@jct.ac.il}
\begin{abstract}
The Aharonov-Bohm effect is traditionally attributed to the effect of the electromagnetic 4-potential $A$, even in regions where both the electric field $\mathbf{E}$
and the magnetic field $\mathbf{B}$ are zero. The AB effect reveals that multiple-valued functions play a crucial role in the description of an electromagnetic field. We argue that the quantity measured by AB experiments is a difference in values of a multiple-valued complex function, which we call a \textit{complex potential} or \textit{pre-potential}. We show that any electromagnetic field can be described by
this pre-potential, and give an explicit expression for the electromagnetic field tensor through this potential.
  The pre-potential is a modification of the two scalar potential functions
introduced by E. T. Whittaker.
\end{abstract}

\submitto{\JPA}
\maketitle

\section{Introduction}
In their seminal paper \cite{AharonovBohm59},
Y. Aharonov and D. Bohm claim that, contrary to the
conclusions of classical mechanics, the electromagnetic
4-potential $A$ affects the motion of an electron beam, even in regions where the electromagnetic field vanishes.
They proposed two kinds of experiments which were successfully
performed later (see, for instance,
\cite{OsakabeMatsudaKawasakiEndoTonomura86} and \cite{van Oudenaarden}).

In  \cite{AharonovBohm59} a scalar function $\mathcal{S}$  such that $\nabla \mathcal{S}=(e\hbar/c)A$ is introduced.
It has been shown that if  $\psi_0$ is the solution of the Schr\"odinger equation in the absence of an EM field, then
the function $\psi=\psi_0e^{-i\mathcal{S}/\hbar}$ is the solution of the equation in the presence
of the field. When the region is multiply connected,
$\mathcal{S}$ is not a single-valued function, and calculating $\mathcal{S}$
by two non-equivalent paths can produce $\mathcal{S}_1-\mathcal{S}_2\neq 0$.
It is argued that in the magnetic AB experiment,
the difference of values of some multivalued function $\mathcal{S}(x)$ is measured.
The function $\mathcal{S}(x)$ is a real-valued function of spatial coordinates and is the logarithm of the phase factor of the corresponding
Schr\"odinger equation.

The ability of the four-potential $A$ to account for the Aharonov-Bohm effect is limited to the case in which the EM field is zero. In this paper, we give a more precise explanation of
the effect by shifting the focus from the four-potential $A$ to a complex-valued multifunction $S$.
Obviously, an everywhere defined \textit{real}-valued function $S$ satisfying $\nabla S=(e\hbar/c)A$ cannot produce a non-trivial field, since in this case the expression $F=\nabla\times A$  vanishes.  But, as we have shown in \cite{FGW}, we can describe an electromagnetic field by a \textit{complex}-valued function $S(x)$, which we call the \textit{scalar complex potential of the electromagnetic field}.
The real and imaginary parts of $S(x)$ are, in fact, the two 'scalar potentials'
introduced by E.~T.~Whittaker in 1904 \cite{Whittaker04}.

 Note that another complex scalar potential was introduced by
H.~S.~Green and E.~Wolf in 1953 \cite{GreenWolf53}. They described the similarity of the expressions for energy and momentum densities between their potential and the wave function. The relation of this potential to the Whittaker one is still unclear. 

Let us describe the function $S(x)$ in more details. Given a moving charge and an observer, we obtain a complex dimensionless scalar.
We then define the scalar complex potential $S$ as the logarithm of this dimensionless scalar.
As in \cite{AharonovBohm59}, our $S$ is not a multi-valued function (it contains complex logarithm). Moreover, we shall see
that the multi-valued nature of $S$
is the \textit{tailor-made} mathematical expression of the Aharonov-Bohm effect. Next, we derive the complex Faraday vector $\bF:=\bE+i\bB$ of an electromagnetic field
from the pre-potential $S$:
\[ F_j={\partial}^\nu (\alpha_j)_\nu ^\lambda{\partial}_\lambda S,\]
where the $\alpha_j$ are Dirac's $\alpha$-matrices.
These matrices $\alpha_k$ are used to insert a
Lorentz invariant conjugation between the gradient and the curl as
they are applied to $S$.
Finally, we presnt a third-order differential equation
expressing the connection between the complex potential $S(x)$ and the field sources.

\section{Definition of the Complex Potential of a moving charge}

Denote by $P=(t,x,y,z)$ a point in space-time at which we want to calculate
the four potential. We call $P$ the observer. Denote by $L$ the
world-line of the charge $q$ generating our electromagnetic field. Let
$Q\in L$ be the unique point of intersection of the past light cone at $P$ with the
world-line $L$ of the charge. We denote the time of the event $Q$ by $\tilde{\tau}$ and refer to this time as the \textit{retarded time} of the potential.
Note that radiation emitted at $Q$ will reach $P$ at time
$t$. Thus, the potential at $P$ will depend only on the position described by the vector $a=\overrightarrow{QP}$  of the charge at the proper time $\tilde{\tau}$. See Figure 1.
\begin{figure}[h!]
  \centering
\scalebox{0.4}{\includegraphics{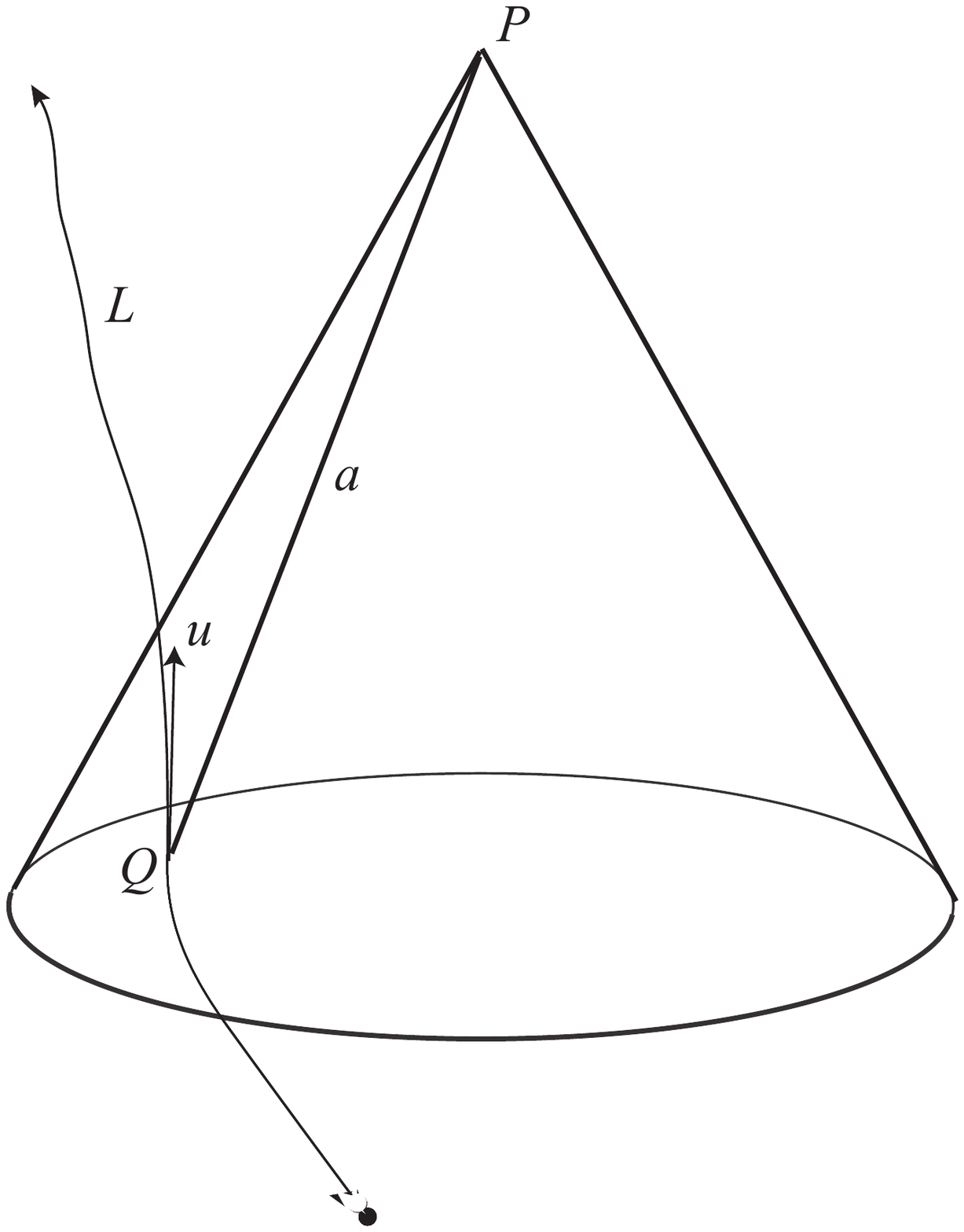}}
  \caption{The four-vectors associated with an observer and a moving charge.}\label{chargePotent}
\end{figure}

Let $K$ be an inertial reference frame in space-time with coordinates $(ct,x,y,z)=x^\mu,$ where $c$ denotes the speed of light.
For the rest of the paper, we will use units in which $c=1$ and omit $c$ from equations. The inner product of two 4-vectors is defined as
\begin{equation}\label{inner prod mink}
  {a}\cdot {b}=\eta _{\mu\nu}a^\mu b^\nu ,\;\;\eta _{\mu\nu}=\mbox{diag}(1,-1,-1,-1).
\end{equation}
The space of 4-vectors with this inner product is Minkowski space-time $M$.
Let $x^\mu$ denote the coordinates of $P$, and let $\tilde{x}^\mu$ be the coordinates of $Q$, the charge at the retarded time.
Introduce a 4-vector $a(x)=\overrightarrow{QP}$. Then
\begin{equation}\label{a def}
 a^\mu(x)=x^\mu-\tilde{x}^\mu \quad \mbox{ and } \quad a^2=a\cdot a=0.
\end{equation}
The vector $a(x)$ is a null (light-like) vector in space-time.

Since $a$ is a null vector, we have
 \begin{equation}\label{null vec relations}
   (a^0+a^3)(a^0-a^3)=(a^1+ia^2)(a^1-ia^2).
 \end{equation}
We may therefore define a dimensionless complex constant
\begin{equation}\label{z in cart}
    \zeta=\frac{a^1-ia^2}{a^0+a^3}=\frac{a^0-a^3}{a^1+ia^2}.
\end{equation}
This constant coincides with the ``single complex parameter" occurring
during the stereographic projection of the celestial sphere to the Agrand plane (see \cite{PenroseRindler} v.1 p.15).

We want the scalar potential to be a function of the dimensionless scalar $\zeta$.
To identify the "right" function, note that the electric force depends on
the distance from the charge as $\frac{1}{r^2}$ and, as explained in the Introduction,
the force is a second derivative of the potential. Hence, the natural candidate for the
scalar potential is a multiple of the logarithm function.
\vspace{.2in}

\noindent\textbf{Definition}\quad We define the \textit{complex potential} or \textit{pre-potential} $S (x)$ at the observer point$x$ of a moving charge $q$ by
\begin{equation}\label{WitPoint_1}
    S(x)=q\ln \zeta=q\ln\frac{a^1(x)-ia^2(x)}{a^0(x)+a^3(x)},
\end{equation}
where $a(x)$ is defined by (\ref{a def}).

\section{The 4-potential and the Faraday vector of the electromagnetic field}

 An electromagnetic field can be defined by an electric field intensity $\mathbf{E}(\mathbf{r},t)$
 and a magnetic field intensity $\mathbf{B}(\mathbf{r},t)$. Equivalently, one can define a complex
3D-vector $\mathbf{F}$, called the Faraday vector, by
\begin{equation}\label{FaradayVector}
 \mathbf{F}=\mathbf{E}+i\mathbf{B}
\end{equation}
in order to represent the electromagnetic field. Since  $\mathbf{E}$ and $\mathbf{B}$  may be expressed as certain
derivatives of the the 4-potential $A=A_\mu$, the Faraday vector $\mathbf{F}$ may also be derived from the 4-potential:
\begin{equation}\label{Fj fromA}
   F_j=2{\partial}^\nu(\rho^ j)^\mu_\nu A_\mu.
\end{equation}
Here the differential operators are $\dmu=\frac{\partial}{\partial x^\mu}$ and
$\Dmu=\eta^{\mu\nu}{\partial}_\nu$. The matrices $(\rho^ j)^\mu_\nu$ are the Majorana-Oppenheimer matrices (see  \cite{Dvoeglasov})
\[
(\rho^1)_\nu^\mu=\frac{1}{2}\left(
\begin{array}{cccc}
0 & 1 & 0 & 0 \\
1 & 0 & 0 & 0 \\
0 & 0 & 0 & -i \\
0 & 0 & i & 0 \\
\end{array}\right),\;
(\rho^2)_\nu^\mu=\frac{1}{2}\left(
\begin{array}{cccc}
0 & 0 & 1 & 0 \\
0 & 0 & 0 & i \\
1 & 0 & 0 & 0 \\
0 & -i & 0 & 0 \\
\end{array}\right),\]
\begin{equation}\label{rho matrices}
 (\rho^3)_\nu^\mu=\frac{1}{2}\left(
\begin{array}{cccc}
0 & 0 & 0 & 1 \\
0 & 0 & -i & 0 \\
0 & i & 0 & 0 \\
1 & 0 & 0 & 0 \\
\end{array}\right)\,.
\end{equation}
which were used and studied also in \cite{FD}.

Direct computation shows that the matrices $\sigma^j=i\rho^j$ obey the commutator relations of the rotation
group $SO(3)$, while the $\rho^j$ matrices  obey the commutator relations of boosts in the Lorentz group, i.e.
\begin{equation}\label{commutationof Kj}
   [\sigma^j,\sigma^k]=-\epsilon^{jk}_l\sigma^l,\;\;[\rho^j,\rho^k]=\epsilon^{jk}_l\sigma^l\;\;
   [\sigma^j,\rho^k]=\epsilon^{jk}_l\rho^l\,.
\end{equation}
As a result, we can use the six matrices $\rho^j,\sigma^j$ as generators of the Lie algebra of the Lorentz group. In addition, the complex conjugates $\bar{\rho}^j$ of
these matrices satisfy the same commutation relations. Thus
$\bar{\rho}^j$ and $\bar{\sigma}^j=-i\bar{\rho}^j$ also generate the Lie algebra of the Lorentz group. Moreover,
\begin{equation}\label{commutation of k kbar}
    [\bar{\rho}^j,\rho^l]=0\,.
\end{equation}

In addition to the above commutation relations, these matrices also satisfy the following anti-commutation relations:
\begin{equation}\label{CAR}
    \{\rho^j,\rho^l\}=\rho^j\rho^l+\rho^l\rho^j=\frac{\delta^{jl}}{2}I,\;\;
    \{\bar{\rho}^j,\bar{\rho}^l\}=\frac{\delta^{jl}}{2}I\,.
\end{equation}

\section{Lorentz group representations in $\cM^4$}

Denote by $\cM^4$ the complex space $\mathbb{C}^4$
endowed with the bilinear $\mathbb{C}$-valued form
$x\cdot y=\eta_{\mu\nu} x^\mu y^\nu$. One possible interpretation of $\cM^4$  is the following.
Let $\psi:M \rightarrow \mathbb{C}$ be the wave function of a zero spin particle.
The gradient operator, describing a generalized momentum, maps space-time into $\cM^4$, since $\nabla\psi\in\mathbb{C}^4$. The bilinear form
on $\cM^4$ is an extension of the inner product on the Minkowski space-time.
Our complex scalar potential $S(x)$ is a function $M \rightarrow \mathbb{C}$ as well.

Denote by  $\pi$ the lift to $\cM^4$ of the fundamental representation of the Lorentz group $L$ on $M$, see
Figure~\ref{pirep}.
\begin{figure}[h!]
\scalebox{0.9}{\begin{picture}(500,120)(-125,-20)
\put(15,70){\makebox(0,0){$\cM^4$}}
\put(40,70){\vector(1,0){120}}
\put(100,80){\makebox(0,0){$\pi$}}
\put(182,70){\makebox(0,0){$\cM^4$}}
\put(17,15){\vector(0,1){40}}
\put(10,35){\makebox(0,0){$\nabla$}}
\put(180,15){\vector(0,1){40}}
\put(173,35){\makebox(0,0){$\nabla$}}
\put(182,1){\makebox(0,0){$M$}}
\put(40,1){\vector(1,0){120}}
\put(100,12){\makebox(0,0){$\Lambda$}}
\put(17,1){\makebox(0,0){$M$}}
\end{picture}}
\caption{The representation $\pi$}\label{pirep}
\end{figure}

 We denote by $\tilde{\pi}$ the representation in $\cM^4$ generated by matrices
$\rho^j$ and $i\rho^j$ (a boost in direction $j$ is given by $\Upsilon ^j=\exp (\rho^j)$), and by $\tilde\pi^*$ the representation in $\cM^4$ generated by matrices
$\bar{\rho}^j$ and $-i\bar{\rho}^j$. Note that the matrices $\rho^j+\bar{\rho}^j$ are generators of boosts in direction $j$. Thus, using (\ref{commutation of k kbar}), the  representation $\pi$ of the Lorentz group $L$ can be
decomposed as
\begin{equation}\label{decomposition of pi boosts}
    \Lambda ^j=\exp (\rho^j+\bar{\rho}^j)=\exp (\rho^j)\exp (\bar{\rho}^j)=\Upsilon ^j\bar{\Upsilon} ^j\,,
\end{equation}
or more generally, for any $g\in L$
\begin{equation}\label{decomposition of pi gen}
    \pi(g)=\tilde{\pi}(g)\tilde\pi^*(g)\,.
\end{equation}

\section{Covariance under the representations in $\cM^4$}

In \cite{FD}  a \textit{complex Faraday tensor} is introduced for the description of an electromagnetic
field, similar to the one introduced by Silberstein \cite{Silberstein}. This tensor is a complex matrix (mixed tensor)
$\mathcal{F}^\beta_\alpha =\sum_{j=1}^3 F_j  (\rho^j)^\beta_\alpha$,
with $F_j$ defined by (\ref{FaradayVector}). We denote its complex conjugate by
$\bar{\mathcal{F}}^\beta_\alpha =\sum_{j=1}^3 \bar{F}_j  (\bar{\rho}^j)^\beta_\alpha\,.$
With this notation, the usual electromagnetic tensor $F^\beta_\alpha $ can be decomposed as
\begin{equation}\label{decomp F}
 F^\beta_\alpha ={\mathcal{F}}^\beta_\alpha+\bar{\mathcal{F}}^\beta_\alpha\,.
\end{equation}
We will now prove two claims.
\vskip0.2cm\noindent
\textbf{Claim 1} \quad The covariance of the tensor $F^\beta_\alpha$ under the representation $\pi$ is equivalent  either to the covariance of ${\mathcal{F}}^\beta_\alpha$ under $\tilde{\pi}$ or, equivalently, to the covariance of $\bar{\mathcal{F}}^\beta_\alpha$ under $\tilde{\pi}^*$.
\begin{proof}
We check the covariance under the boost $ \Lambda ^j$, defined by (\ref{decomposition of pi boosts}), in the direction $j$. Under this transformation, from (\ref{decomp F}) we have
\[ F^{\beta'}_{\alpha'}=(\Lambda ^j)^{-1}F^\beta_\alpha\Lambda ^j=(\Upsilon ^j\bar{\Upsilon} ^j)^{-1}(
{\mathcal{F}}+\bar{\mathcal{F}})\Upsilon ^j\bar{\Upsilon} ^j.\]
Using (\ref{commutation of k kbar}), we get $[\Upsilon ^j,\bar{\rho}^l]=[\Upsilon ^j,\bar{\Upsilon}^l]=
[\Upsilon ^j,\bar{\mathcal{F}}]=[\bar{\Upsilon} ^j,\rho^l]=[\bar{\Upsilon} ^j,\mathcal{F}]=0$. Hence, the above equation can be rewritten as
\[(\Lambda ^j)^{-1}F^\beta_\alpha\Lambda ^j=(\Upsilon ^j)^{-1}{\mathcal{F}}\Upsilon ^j+(\bar{\Upsilon} ^j)^{-1}
\bar{\mathcal{F}}\bar{\Upsilon} ^j.\]
This proves Claim 1 for boosts. Similarly, one can establish covariance under action of an arbitrary element of the group $L$.
\end{proof}

\vskip0.2cm\noindent\textbf{Claim 2} \quad The dimensionless constant $\zeta$ defined by (\ref{z in cart}) and the complex potential $S(x)$ defied by (\ref{WitPoint_1}) are covariant under the representation $\tilde{\pi}$.
\begin{proof}
Note that from (\ref{CAR}), it follows that $\Upsilon ^j=\exp(\rho^j\psi)=
\cosh(\psi/2)I+\sinh(\psi/2)2\rho^j.$ Thus, if we apply, for example, $\Upsilon ^1$ on the vector $a$, we get
\[\Upsilon ^1 a=\cosh(\psi/2)(a^0,a^1,a^2,a^3)+\sinh(\psi/2)(a^1,a^0,-ia^3,ia^2).\]
So, applying this transformation to  $\zeta$ and using the identity in (\ref{z in cart}), we get
\[\Upsilon ^1 (\zeta)=\frac{\cosh(\psi/2)(a^1-ia^2)+\sinh(\psi/2)(a^0-a^3)}
{\cosh(\psi/2)(a^0+a^3)+\sinh(\psi/2)(a^1+ia^2)}=\]
\[\frac{a^1-ia^2}{a^0+a^3}\cdot
\frac{1+\tanh(\psi/2)(a^0-a^3)/(a^1-ia^2)}
{1+\tanh(\psi/2)(a^1+ia^2)/(a^0+a^3)}=\zeta.\]
This proves Claim 2.
\end{proof}

\section{The Faraday vector and the complex potential of a uniformly moving charge}

To define the 4-potential $A$ and consequently the field strength $\mathbf{F}$, we need a new
operation on  $\cM^4$. This operation acts by multiplication with the matrix $C=2\bar\rho ^3$, namely,
$a^\mu\mapsto C^\mu_\nu a^\nu$. Since the square of this operation is the identity, we call it the \textit{conjugation}. From (\ref{commutation of k kbar}) it follows that this conjugation is covariant under the representation $\tilde{\pi}$.

Define the complex 4-potential $A$ as the conjugate of the gradient of $S$, i.e.
\[A_\mu=C_\mu^\lambda{\partial}_\lambda S.\]
The Faraday vector $\mathbf{F}$ can be derived from the complex potential by use of (\ref{Fj fromA}):
\begin{equation}\label{F from S}
 F_j={\partial}^\nu (\rho^j)^{\mu}_\nu C_\mu^\lambda{\partial}_\lambda S.
\end{equation}

This gives explicit formulas for each component of the Faraday vector $F$:
\begin{equation}\label{F1 from S}
   F_1=S_{,13}+iS_{,02},\;\;
F_2=S_{,23}-iS_{,01}
\end{equation}
and
\begin{equation}\label{F3 from S}
  F_3=\frac{1}{2} (S_{,00}-S_{,11}-S_{,22}+S_{,33}) \,.
\end{equation}
By the above two claims, equation (\ref{F from S}) is covariant under the representation
$\tilde{\pi}$. Hence, we will compare this formula with known results only in the case of a rest charge at the origin.

Consider a rest charge at the origin.
In this case, the world-line of the charge is $L=(t,0,0,0)$.
From definition (\ref{a def}), we get $a=(|x|,x^1,x^2,x^3)$, where
$|x|=\sqrt{(x^1)^2+(x^2)^2+(x^3)^2}$. Thus,
\[S(x)=q\ln\frac{x^1-ix^2}{|x|+x^3}\,.\]
Let $\varrho =(x^1)^2+(x^2)^2$. Then, since $\frac{\partial}{\partial x^j}|x|=\frac{x^j}{|x|}$, we obtain $S_{,0}=0$,
\[S_{,1}=q\left(\frac{1}{x^1-ix^2}-\frac{x^1/|x|}{|x|+x^3}
\right)=\frac{q}{\varrho}\left(x^1+ix^2-\frac{(|x|-x^3)x^1}{|x|}\right)=
\frac{q}{\varrho}\left(\frac{x^1x^3}{|x|}+ix^2\right),\]
\[S_{,2}=\frac{q}{\varrho}\left(\frac{x^2x^3}{|x|}-ix^1\right),\;\; S_{,3}=-\frac{q}{|x|}\,.\]

Then, from (\ref{F1 from S}),  we obtain
\[ F_1=S_{,13}+iS_{,02}=\frac{\partial}{\partial x^1}S_{,3}=-\frac{\partial}{\partial x^1}
\frac{q}{|x|}=\frac{q x^1}{|x|^3}\]
and
\[ F_2=S_{,23}-iS_{,01}=\frac{\partial}{\partial x^2}S_{,3}=-\frac{\partial}{\partial x^2}
\frac{q}{|x|}=\frac{q x^2}{|x|^3}\,.\]
To calculate $F_3$ using (\ref{F3 from S}), we first calculate
\[S_{,11}=\frac{\partial}{\partial x^1}\frac{q}{\varrho}
\left(\frac{x^1x^3}{|x|}+ix^2\right)=-\frac{2qx^1}{\varrho}\left(\frac{x^1x^3}{|x|}+ix^2\right)+
\frac{q}{\varrho^2}\frac{x^3|x|-x^1x^3\frac{x^1}{|x|}}{|x|^2},\]
\[S_{,22}=\frac{\partial}{\partial x^2}\frac{q}{\varrho}
\left(\frac{x^2x^3}{|x|}-ix^1\right)=-\frac{2qx^2}{\varrho}\left(\frac{x^2x^3}{|x|}-ix^1\right)+
\frac{q}{\varrho^2}\frac{x^3|x|-x^2x^3\frac{x^2}{|x|}}{|x|^2},\]
implying that
\[S_{,11}+S_{,22}=-\frac{q x^3}{|x|^3}\,.\]
Since $S_{,00}=0$ and $S_{,33}=\frac{q x^3}{|x|^3}$, equation (\ref{F3 from S}) yields
\[F_3=\frac{q x^3}{|x|^3}.\]
This coincides with the usual formula for the electric force of a rest charge.

Note that in this case, our complex potential satisfies the wave equation
\begin{equation}\label{wave}
    \square  S  =S_{,00}-S_{,11}-S_{,22}-S_{,33}=S_{,00}-\nabla ^2S=0\,.
\end{equation}
Since the d'Alembertian is  covariant, the wave equation holds for a field generated by any uniformly moving charge
and more generally for any EM field. 

For a  charge $q$  moving uniformly with 4-velocity $u$,  the Faraday vector $\mathbf{F}$ can be calculated by
 \begin{equation}\label{Fjfinal}
    F_j=q\frac{a_\mu ({\rho}^j)_\nu^\mu u^\nu}{(a \cdot u)^3}\,,
\end{equation}
where $({\rho}^j)_\nu^\mu$ are defined by (\ref{rho matrices}). From the above calculations for a rest charge at the origin the equation (\ref{Fjfinal}) holds (in this case $u=(1,0,0,0)$). Since this formula is covariant,
it also holds in the case of a uniformly moving charge.
Equation (\ref{Fjfinal}) coincides  with the usual formula for the field of a
moving charge (see, for example, \cite{Jackson} p. 573). 

\section{The scalar potential for an electromagnetic field}

Any electromagnetic field is generated by a collection of moving charges. We may
assume that charges close to each other move with velocities that do not vary significantly.
The sources of the electromagnetic field may be represented by the charge densities $\sigma (y)$
on the space-time 4-vector $y$. We assume that the potential depends additively on the charges generating the field.
Thus, the scalar complex potential of the electromagnetic field is given by
\begin{equation}\label{scalPotGenera}
    S(x) =\int\limits_{K^-(x)}\ln\left(\frac{a^1-ia^2}{a^0+a^3}\right)\sigma(x+a)da,
\end{equation}
where $K^-(x)$ denotes the backward light-cone at $x$.

 The operators $\alpha_j:=\rho^j C$ occurring in  (\ref{F from S}) satisfy the canonical anti-commutation relations similar to (\ref{CAR}) of Dirac's $\alpha$-matrices.
 Therefore, for any complex potential $S$,  equation (\ref{F from S}) can be rewritten as
 \begin{equation}\label{F from S alfa}
F_j={\partial}^\nu (\alpha_j)_\nu ^\lambda{\partial}_\lambda S\,.
\end{equation}
In the Newman-Penrose basis of $\cM^4$ (also known as Bondi tetrad, see \cite{FGW}), the matrices $\alpha_j$ take the usual form
$\left(                \begin{array}{cc}
                  \sigma _j & 0 \\
                  0 & -\sigma _j \\
                \end{array}
              \right)$
   where $\sigma _j$ are the Pauli matrices.

              Note that the matrices $\rho^j$, which define the representation $\tilde{\pi}$,
also satisfy the canonical anti-commutation relations (\ref{CAR}). However, they  cannot be completed by a $\beta$ matrix, needed for the Dirac equation.
The representation  $\tilde{\pi}$ is a representation of pairs of
spinors, while the representation  $\tilde{\pi}^*$ is
a representation of pairs of dotted spinors.

If the electromagnetic field sources are $J_\mu=(\rho,-j^1,-j^2,-j^3)$, it can be shown \cite{FMaxwell}
that the Maxwell equations become
\[      \partial_\alpha (\nabla ^2 S)=C_\alpha^\beta J_\beta,\]
for $\alpha=0,1,2,3$, added with the wave equation
\[ \square S=0\,.\]

\section{Discussion}

We introduced a new description of an electromagnetic field by a complex-valued function $S(x)$ (pre-potential) on
Minkowski space-time. The advantages of this approach are as follows:
\begin{itemize}
  \item The multiple-valued nature of the pre-potential is a natural expression of the Aharonov-Bohm effect.
  \item Our approach reduces the degrees of freedom from 4 to 2 in the description of an electromagnetic field.
  \item It reveals a new connection between the Dirac equation and classical electrodynamics.
  \item We obtain a new complex Lorentz invariant $\zeta$ associated with any null-vector.
  \item Our approach reveals a new connection between the fundamental and spinor representations of the Lorentz group.
\end{itemize}

Our future steps are:\begin{itemize}
                       \item Incorporate the pre-potential into the Dirac and Schr\"odinger equations.
                       \item Understand the effect of the electromagnetic field on the solutions of these equations through the pre-potential.
                       \item Derive the formulae for the pre-potential for standard sources of an electromagnetic field.
                     \end{itemize}

\ack

We would like to thank  T. Scarr for editorial comments.

\end{document}